\documentclass[prd,amsfonts,amssymb,amsmath,showpacs,nofootinbib,twocolumn]{revtex4-1}
\usepackage[dvips]{graphicx}

\usepackage[normalem]{ulem} 

\begin{document}

\title{Horizon spectroscopy in and beyond general relativity}

\author{Jozef Sk\'akala}
\email{email: jozef@iisertvm.ac.in}

\author{S. Shankaranarayanan} 
\email{e-mail: shanki@iisertvm.ac.in} 
\affiliation{School of Physics, Indian Institute of Science, Education and Research (IISER-TVM), Trivandrum 695016, India}

\begin{abstract}
In this work we generalize the results for the entropy spectra typically derived for black holes 
in general relativity to a generic horizon within the spherically symmetric (asymptotically flat and 
non-flat) space-times of more general theories of gravity. We use all the standard --- Bekenstein's universal lower 
bound on the entropy transition, the highly damped quasi-normal modes and reduced phase-space quantization --- approaches to 
derive the spectra. In particular, the three approaches show that the Bekenstein-like spectra for the 
horizon entropy is a \emph{robust} result. Our results confirm the suggestion made relatively 
recently by an independent fourth argument by Kothawala et al \cite{Paddy1}.
\end{abstract}

\maketitle

\section{Introduction}

The idea that the area/entropy of the black-hole horizon in general relativity (GR)
is quantized originates from the work of Bekenstein \cite{Bekenstein1}. Bekenstein's 
arguments suggest that the semi-classical black hole (BH) entropy spectrum is given by,
(let us keep for the rest of the paper Boltzmann constant $k=1$ and also $G=\hbar=c=1$):
\begin{equation}\label{entropy}
S=2\pi\gamma\cdot n, ~~~\gamma\in O(1).
\end{equation}
Following Bekenstein, the BH entropy spectrum (\emph{often}
with $\gamma=1$) was re-derived in many different ways \cite{Bekenstein2,
  Kunstatter1, Kunstatter2, Kunstatter3, Medved, Makela1, Makela2, Simon, Hod, Maggiore, KunstatterQNM}. These
derivations can broadly be classified into three different types of 
arguments: 
\begin{enumerate}
 \item  Bekenstein's original argument about
the universality (independence on BH parameters) of the lower bound on
the BH entropy/area transition \cite{Bekenstein1}. 
\item Argument via asymptotically highly damped BH quasi-normal modes \cite{Hod, Maggiore,
    KunstatterQNM}.
\item Direct quantization techniques typically applied to a reduced phase space of
the black hole parameters \cite{Bekenstein2, Kunstatter1,
    Kunstatter2, Kunstatter3, Medved, Makela1, Makela2, Simon}.     
\end{enumerate}
To these three arguments a fourth independent argument was put forward 
relatively recently by Kothawala et al \cite{Paddy1}. The authors argument relies 
on the use of effective action of a class of observers constrained in a spacetime 
region with a horizon. The authors show that the semiclassical propagator that satisfies
the WKB approximation is consistent if the entropy fulfils (at least for Lanczos-Lovelock 
theories) quantization condition of the form \eqref{entropy} with $\gamma=1$. The argument by \cite{Paddy1}, being 
fairly general, consequently suggests that the result for the BH entropy 
spectra in general relativity may be generalized in two main directions: 
First, it may hold for entropy of an arbitrary space-time horizon (observer dependent, 
or not). Secondly, it may hold for Wald entropy in much more general theories than 
the general theory of relativity (at least within Lanczos-Lovelock theories). 

The purpose of this work is to use the three standard arguments mentioned above 
(which were originally used to derive the BH entropy spectra within general relativity), 
to confirm or infirm  the conclusions of Kothawala et al \cite{Paddy1}. (This work also
continues previous work done by one of the authors in \cite{Skakala1,
Skakala2}.)

The main conclusion of this work is that, indeed, for \emph{general horizons} in 
spherically symmetric sector within \emph{more general theories} than general relativity
(GR) one has completely the same
evidence for the entropy spectra of the Bekenstein type, as one had in the particular case of BH horizon in GR. 
This means, the quantization of entropy is a very \emph{robust} result, such that
is related to the general thermodynamics of horizons, rather than only
an artefact of the black hole theory in GR. (However it might be not
surprising that it was first time discovered in the black hole
context.)  The basic conclusion of this paper therefore confirms the suggestion of Ref. \cite{Paddy1}.  
Our analysis, as in Ref. \cite{Paddy1}, shows that in such theories of gravity where proportionality between the 
horizon area and the horizon entropy does \emph{not} hold, one has to expect 
\emph{horizon entropy, not the horizon area}, to have an equispaced spectrum. 
Furthermore our conclusion perfectly matches with the longer term development in 
the field of spacetime thermodynamics, where first the concept of temperature \cite{Unruh,
  Gibbons, Peltola, Paddy2} and subsequently also concept of entropy
\cite{Paddy2, Jacobson, Bianchi} were generalized from black holes to the
general horizons and also to more general gravity theories than GR
\cite{Wald}.

The rest of the paper is organized as follows: In Sec. (II), we discuss 
the generalization of Bekenstein's argument to general horizons in spherically symmetric solutions of generic theories of gravity. In Sec. (III), 
we generalize the asymptotic QNM analysis to generic theories of gravity.
In Sec. (IV), we discuss the detailed procedure of generalization of the constrained phase space 
approach to general horizons in generalized theories of gravity. We would like 
to point that the detailed analysis in Sec. (IV) generalizes to Lovelock theories 
of gravity and conclude in Sec. (V). 
As mentioned earlier, we set Boltzmann constant $k=1$ and also $G=\hbar=c=1$) and the metric convention 
is $(-,+,+,+)$. 

\section{Argument 1: Bekenstein's lower bound on the area / entropy transition}

Let us consider a $D$ dimensional static spacetime with maximally symmetric $D-2$ dimensional subspace, which can be expressed in suitable fixed coordinates as:
\begin{equation}\label{static}
ds^{2}=-f(r)dt^{2}+f(r)^{-1}dr^{2}+g^{(D-2)}_{ij\bot}dx^{i}dx^{j}.
\end{equation}
Here $i,j=2, \dots ,D-1$. For example, the $(D - 2)$ subspace of a 
general static spherically symmetric line element \eqref{static} 
can be expressed as
\[g^{(D-2)}_{ij\bot}dx^{i}dx^{j}=h^{2}(r)d\Omega_{D-2}^{2}~.\] 
Note that the line element \eqref{static} includes also Rindler spacetime and therefore the method used in this section applies also to the Rindler horizon. However both the quasi-normal mode analysis and the reduced phase space quantization cannot be applied to Rindler spacetime: The quasi-normal mode analysis does not apply to apparent horizons, such as Rindler horizon, and the reduced phase space approach requires the maximally symmetric subspace to be sphere, which Rindler spacetime does not fulfil. 

Without loss of generality, in the static region of the metric \eqref{static}, 
we can assume $f(r)>0$ and let us further consider that the line element has at
least one horizon defined by $f(r_{H})=0$.
Considering the line element \eqref{static} let us consider a gravity
theory where one can write down a quasi-local version of the first law
of thermodynamics (this can be done for a fairly generic theory
\cite{Paddy2}):
\begin{equation}\label{firstlaw}
T_{H}\delta S_{H}=\int_{H} T_{ab}\xi^{a}d\Sigma^{b}.
\end{equation}
($\xi^{a}$ is suitably normalized time-like Killing vector, $T_{H}$ is
temperature of the horizon and $S_{H}$ is Wald entropy of the horizon. 
The integral on the right side is the energy flow through the
Killing horizon. Note also that as a matter of principle we want our
results and methods to be always \emph{quasi-local}, as none of the
physically relevant results should depend on the global structure of
spacetime.) 

Let us consider a point particle moving along the spacetime geodesics passing 
through the horizon.  The point particle is required to contribute to the stress-tensor 
through its mass and does not change any other additional parameters, 
(such as the electromagnetic charge). Under this scenario, 
the change in the entropy will only correspond to the change in the mass parameter. 


It is easy to observe that the entropy will have a minimum change when
the particle is dropped down the horizon radially. If the particle
with a rest mass $m$ has somewhere above the horizon a classical
turning point $r_{p}$, the Killing energy flow through the horizon
will correspond to: 
\[E=m\sqrt{f(r_{p})}.\]

If the classical turning point of the point particle is arbitrarily close 
to the horizon, the change in entropy is expected to be arbitrarily
small. Analogous to Bekenstein's \cite{Bekenstein1} argument, this suggests that 
the Wald entropy is a classical adiabatic invariant.  Due to Ehrenfest principle \cite{Bekenstein1,
Bekenstein2, Bekenstein3}, the Wald entropy should have semi-classically discrete
spectrum.

According to Bekenstein \cite{Bekenstein1}, the quantum corrections can be included by 
attributing the particle an effective size. This places constraints on the entropy transitions.
Following Bekenstein, the particle's center of mass should move along the geodesics. 
To minimalize the change of entropy, let us place the center of mass of the particle to such a distance from the horizon that 
corresponds to a proper radius
of the particle $b$. Let us approximate the line element
\eqref{static} near the horizon by a Rindler line element using
\[f(r)=f_{,r}(r_{H})\{r-r_{H}\}+O(r^{2})=2\kappa\cdot\{r-r_{H}\}+O(r^{2}),\]
with $\kappa$ being the surface gravity of the horizon. The line element 
\eqref{static} turns (by redefining $x\doteq r-r_{H}$) into the Rindler line element:

\begin{equation}\label{Rindler}
-2\kappa x\cdot dt^{2}+(2\kappa x)^{-1}dx^{2}+dL^{2}_{\bot}.
\end{equation} 

The radial position of the particle's center of mass at the classical turning point is
related to the proper radius of the particle as:
\[b=\int_{0}^{\delta}f^{-1/2}(x)dx=\int_{0}^{\delta}\frac{dx}{\sqrt{2\kappa x}}
=\sqrt{\frac{2\delta}{\kappa}}.\]
Here $\delta$ is the radial position of the particle's center of mass. The Killing 
energy that had flown under the horizon is given as 
\[E=m\sqrt{2\kappa\delta}=|\kappa|\cdot mb.\]
Using the fact that the horizon temperature for the line element \eqref{Rindler} is  given by
\[T_{H}=\frac{|\kappa|}{2\pi},\] 
the first law of thermodynamics \eqref{firstlaw} gives: 
\[\frac{|\kappa|}{2\pi}\delta S_{min}=|\kappa|\cdot mb~~~~\to~~~~~\delta S_{min}=2\pi mb \, ,\]
where $\delta S_{min}$ is the minimal increase of Wald entropy due to
the absorption of a particle with parameters $(m,b)$.  Then the
Bekenstein's original argument extended to Wald's entropy goes as follows: 
One can not choose the parameters $(m, b)$ of the particle arbitrarily, but 
the particle's proper radius $b$ has to be bounded either by the reduced Compton
wavelength of the particle (Uncertainty principle), or by
the Schwarzschild radius, whichever is larger. In Planck units, the reduced Compton
wavelength is larger for $m<2^{-1/2}$ and the Schwarzschild radius is larger for $m>2^{-1/2}$. 
If the Compton wavelength is larger, then $m\geq b^{-1}$, therefore $m \, b\geq 1$ and
then $\delta S_{min}$ is lower bounded (in Planck units) by
\[\delta S_{min}\geq 2\pi.\]
In the case where the Schwarzschild radius is larger, then $b\geq 2m$, but then 
\[bm\geq 2m^{2}\geq 1,\]
and one obtains exactly the same lower bound on the entropy transition $\delta S_{min}\geq 2\pi$. 

Therefore we see that Wald entropy transition has a lower-bound whose value is $ 2 \pi$.  
It is important to note that the result is universal \emph{for any horizon}, being \emph{independent} on the
spacetime near-horizon parameters (the horizon surface
gravity) and the coupling constants of the modified gravity models. 
The requirement on the gravity theory is that it fulfils the first
law of thermodynamics in the form \eqref{firstlaw}. 

According to Bekenstein's original argument, the existence of  the 
\emph{universal} lower bound of the BH area/entropy transition arises 
due to the fact that the BH horizon entropy/area is equispaced as given by 
Eq. \eqref{entropy}. 
Following the same reasoning the result in this section applied to 
any quasi-local horizon of spherically symmetric static spacetime in any \emph{generic} theory of gravity implies 
that that the entropy is equispaced and is given by formula \eqref{entropy}. 
($\gamma$ parameter reflects some ambiguities in the derivation of the 
lower bound.) 

(We would like to mention that the Bekenstein area / entropy lower bound 
was derived within GR for the Kerr-deSitter spacetime and the deSitter horizon 
in\footnote{We would like to thank Jacob Bekenstein for pointing this
reference to us.} \cite{Schiffer}.)  
  
\section{Argument 2: The highly damped quasi-normal modes}

The second argument used to derive the black hole entropy / area
spectra comes originally from the work of Hod and Bekenstein \cite{Hod, Bekenstein2}. It
was suggested that due to the Bohr's correspondence principle ``the
transition frequencies at high quantum numbers equate the classical
oscillation frequencies'' one could possibly identify the transition
frequency between different BH states (in the semi-classical limit)
with the BH quasi-normal mode frequencies. It was observed in
Ref.~\cite{Hod} that since the transitions at high quantum numbers are
supposed to have relaxation times close to zero, the relevant
quasi-normal frequencies are the ones in the limit of high
damping. (We would refer the readers to Refs. \cite{Nollert, Cardoso} 
for comprehensive reviews on quasi-normal modes.) 

The original conjecture showed some difficulties (for example, see 
Ref. \cite{Zhidenko,Shanki1,Shanki2}). However, difficulties in Hod's conjecture was overcome by Maggiore's 
conjecture\cite{Maggiore} to link the asymptotic highly
damped frequencies to the transition mass (in Planck units) as:
\[\Delta M=\lim_{n\to\infty}\Delta_{(n,n-1)}\sqrt{\omega_{nR}^{2}+\omega^{2}_{nI}}=\lim_{n\to\infty}
\Delta_{(n,n-1)}\omega_{nI},\] 
where $\Delta_{(n,n-1)}$ refers to the difference between two adjacent levels. 
(Here we write the quasi-normal frequencies as $\omega=\omega_{R}+i\omega_{I}$. Also, the reason for the
upper equality between the limits is that for all the relevant cases the real part of the
frequency is bounded, while the imaginary part is unbounded.) (For the
detailed reasoning why the upper connection is made see again the original paper
\cite{Maggiore}. Alternatively some more detailed analysis of the
conjectures linking the asymptotic quasi-normal modes and the transition
masses was offered recently by one of the authors \cite{Skakala2}.) In the rest 
of this section, we will show that Maggiore's modification can be used to support the main
conclusions of this paper.

In Ref. \cite{Shanki1, Shanki2} one of the authors derived
general transcendental formulas for the asymptotic quasi-normal
frequencies of tensor perturbations for a generic single horizon black hole in asymptotically flat, de Sitter and Anti-de Sitter static spherically 
symmetric spacetime. In particular the analysis of \cite{Shanki1, Shanki2} depends only on the 
properties of the metric near the horizon and singularity, and is independent 
on the form of gravity theory and matter contribution. Let us put aside 
the asymptotically anti-deSitter spacetime, where it is for specific 
reasons (absence of asymptotically highly damped frequencies) unsuitable 
to use Maggiore's conjecture.

Let us first proceed for the generic asymptotically flat spherically
symmetric static spacetime with a single horizon. For such a spacetime
it was shown in Ref. \cite{Shanki1} that:
\begin{equation}\label{freq}
\omega_{n}=\hbox{(offset)}+in\cdot\kappa +O(n^{-1/2}),
\end{equation}
where ``offset'' is some complex number and $\kappa$ is the surface gravity of the horizon. 
This universal spacing of quasi-normal
modes in spherically symmetric static spacetimes can be confirmed via
the intuition obtained through the Born approximation \cite{Visser,
  Paddy3}, or approximations by the analytically solvable models
\cite{Skakala3}. Using \eqref{freq}, the following relation holds:
\begin{eqnarray}\label{asymptotic}
\lim_{n\to\infty}\Delta_{(n,n-1)}\sqrt{\omega_{nR}^{2}+\omega^{2}_{nI}}=~~~~~~~~~~~\\
\lim_{n\to\infty}
\Delta_{(n,n-1)}\omega_{nI}=\kappa=2\pi T_{H},\nonumber
\end{eqnarray}
where $T_{H}$ is the horizon's temperature. One can further plug the result \eqref{asymptotic} to
Maggiore's conjecture and derive for the mass quantum the following:
\[\delta M=2\pi T_{H}.\]
Assuming again that the spherically symmetric static spacetime is a
solution of a theory in which the quasi-local first law of
thermodynamics \eqref{firstlaw} holds\footnote{Similarly to the previous section, perturbations considered here do not carry any additional properties beyond energy (like electromagnetic charge).}, one automatically obtains for
the entropy :
\[T_{H}\delta S_{H}=\delta M=2\pi T_{H}, ~~~~~\to~~~~~\delta S_{H}=2\pi.\]

One can extend the above result to the \emph{generic} spherically
symmetric static two horizon asymptotically deSitter spacetimes. In
Ref. \cite{Shanki2} it was shown that the formula for the asymptotic
frequencies can be for the generic case put in the form which was
later on reasonably general grounds analysed in \cite{Skakala4}. The
formula for the frequencies can be written as:
\begin{equation}\label{QNMs}
\sum_{i=1}^{K}\left\{A_{i}\exp\left(B_{i1}\frac{\omega}{T_{1}}+B_{i2}\frac{\omega}{T_{2}}\right)\right\}=0.
\end{equation} 
Here $K$ and $A_{i}$ are numbers of no particular importance from the
point of what we want to show, on the other hand it is important to
mention that $B_{i1}, B_{i2}$ are integers. Further by $T_{1}, T_{2}$
we mean temperatures of the two horizons (the BH horizon and the
deSitter cosmological horizon). It can be shown \cite{Skakala4} that
if the ratio of the two temperatures is a rational number, the
solutions of \eqref{QNMs} split in a \emph{finite} number of families
labeled by $a$ of the form:
\[\hbox{(offset)}_{a}+i n\cdot 2\pi\cdot\hbox{lcm}(T_{1},T_{2}).\]
Here again the ``offset'' is some complex number depending this time on the family and
by ``lcm'' we mean the least common multiple of the two temperatures
in question, therefore:
\[\hbox{lcm}(T_{1},T_{2})=p_{1}T_{1}=p_{2}T_{2},\]
where $p_{1}, p_{2}$ are relatively prime integers.

Let us further employ the reasoning from \cite{Skakala1} and assume
what would happen from the point of Maggiore's conjecture if we
assumed both semi-classical entropy spectra of the both horizons to be
in Planck units $2\pi\gamma n$. Let us, for example, imagine that a quantum of
mass appears from the white hole horizon and disappears eventually
behind the cosmological horizon. Due to the first law of
thermodynamics \eqref{firstlaw}, if the Killing energy between the
horizons remains eventually unchanged the quantum of mass has to
fulfil:
\[\delta M=-T_{1}\delta S_{1}=T_{2}\delta S_{2}.\]
If the spectra of $S_{1}, S_{2}$ are both of the form $2\pi\gamma n$ then
the transition in entropy of the horizons can be of only the form:
\[\delta S_{1,2}=2\pi\gamma m_{1,2},\]
where $m_{1,2}$ are some integers. Therefore the mass quantum has to
fulfil:
\[\delta M=-T_{1}\cdot 2\pi\gamma m_{1}=T_{2}\cdot 2\pi\gamma m_{2},\]
but if we want the quantum to be as small as possible, such that it is
consistent with the previous condition then necessarily:
\[\delta M=2\pi\gamma \cdot\hbox{lcm}(T_{1},T_{2}).\]
Maggiore's conjecture then implies that:
\[\lim_{n\to\infty} \Delta_{(n,n-1)}\omega_{nI}=2\pi\gamma\cdot\hbox{lcm}(T_{1},T_{2}),\]
which is precisely the case we observe for each of the families if $\gamma=1$. (One
has to therefore attribute the genuine physical meaning only to the
families of the frequencies. Also let us mention that the physics of
the situation considered, together with the universal $2\pi n$
entropy spectra has as a consequence the rational ratio of
temperatures of the two horizons. However, one does not want to
speculate if this has any real physical meaning, but notice in this
sense a complementary result obtained in Ref. \cite{Paddy4}.)

\section{Argument 3: Quantization of generic horizons in geometrodynamics of spherically symmetric regions of spacetimes}

The third argument for the spectra \eqref{entropy} is obtained
through a direct quantization of a constrained phase space corresponding to 
the spacetime variables. There is more than one way on how to derive 
the form of entropy spectra within this approach.  We consider the approach of \cite{Kunstatter1, Kunstatter2, Kunstatter3} to be the most straightforward and based on least assumptions. We will further use the 2D effective dilaton 
geometrodynamical approach \cite{Cavaglia1, Cavaglia2, Kunstatter4, Kunstatter5} to apply 
it to general horizons in generalized gravity theories. The 2D dilaton gravity
is chosen because it describes dimensionally reduced spherically
symmetric sector of $D$ dimensional general relativity with the
cosmological constant, and near horizon-limit of the general 
$D$ dimensional Lovelock gravity. After introducing the
geometrodynamics of the 2D dilaton gravity with the EM field, we
review the suggestion by \cite{Kunstatter1, Kunstatter2, Kunstatter3}
of how to quantize the Euclidean sector of the theory. 

Let us mention that the authors of \cite{Kunstatter1, Kunstatter2, Kunstatter3} claim 
that their quantization results are fairly general within black holes in 
generalized theories, however we suggest that there is no reason to be 
constrained by black holes: the result relates to general spherically symmetric horizons.
Therefore, as we will see, the quantization techniques could be used in the 
same way to derive the entropy spectra (for example) of the deSitter cosmological horizon, 
or of the inner Cauchy horizon of the Reissner-Nordstr\"om black hole. In this section, we 
further show that the technique can be used to derive the entropy spectrum for the 
spherically symmetric sector in GR in arbitrary dimension with
cosmological constant, and, with some approximation, it further
generalizes to Wald entropy within general $D$ dimensional spherically
symmetric Lovelock gravity.

We will describe in detail the Hamiltonian formulation of geometrodynamics 
of a 2D dilaton theory coupled to the Maxwell field, despite of the fact 
that most of the different pieces of the formalism used here can be found 
in relatively broad literature on the subject \cite{Cavaglia1, Kuchar, Kunstatter4, Kunstatter5, Louko1,
Louko2}. In Ref. \cite{Kunstatter4, Kunstatter5}, the authors have looked at a general 2D dilaton 
gravity with EM fields, however, this section offers a more comprehensive analysis necessary 
to reduce the theory to effective 1-dimensional action. At the end of the detailed analysis we 
offer a relatively broad discussion that 
reflects the shift in the viewpoint that we suggest.

\subsection{Geometrodynamics of the 2D dilaton gravity with the E-M field}

As mentioned earlier, we will consider a general version of a dilaton-EM field action (some of the
notation and factors match \cite{Kunstatter4}):
\begin{equation}\label{Gaction}
S_{_{2D}} = \frac{1}{2}\int \sqrt{-g}\left[\phi R^{(2)}+V(\phi)-\frac{W(\phi)}{2}F^{(2)\mu\nu}F^{(2)}_{\mu\nu}\right].
\end{equation}
and parametrize the 2D metric as:
\[ds^2 = g_{\mu\nu}dx^{\mu}dx^{\nu}=-N^{2}dt^{2}+B^{2}(dx+N^{x}dt)^{2}.\]

One can express the action \eqref{Gaction} as:
\begin{eqnarray}\label{Gaction2}
S_{_{2D}} &=&\int dt dx\cdot \left[P_{B}\dot B+P_{\phi}\dot\phi+P_{A_{x}}\dot A_{x}-NH\right.~~~~~~~\nonumber\\
& & \qquad \qquad \left.- N^{x}H_{x} -\tilde A_{t}H_{A}\right],~~~~
\end{eqnarray}
where $A_{\mu}$ is the electromagnetic potential,
\[P_{B}=\frac{-\dot\phi+N^{x}\phi'}{N},\]
\[P_{\phi}=\frac{(N^{x}B)'-\dot B}{N},\]
\[P_{A_{x}}=\frac{W(\phi)(\dot A_{x}-A_{t}')}{BN},\]
\[\tilde A_{t}=A_{t}-N^{x}A_{x},\] 
and the constraint variables read as:
\[H_{t}=-P_{B}P_{\phi}-\frac{1}{2}BV(\phi)+\frac{\phi''}{B}-\frac{B'\phi'}{B^{2}}+\frac{BP_{A_{x}}^{2}}{2W(\phi)},\]
\[H_{x}=P_{\phi}\phi'-\Lambda P_{\Lambda}'-A_{x}P_{A_{x}}',\]
\[H_{A}=-P_{A_{x}}'.\]

Dot denotes $t$-derivative and prime denotes $x$-derivative.

In order to define the variational problem, one has to
supplement it with the correct fall-off conditions of the functions at
the boundaries. One obtains suitable boundary conditions by simply
updating the fall-off conditions given by \cite{Cavaglia1}.  We will
further follow the ideas of \cite{Kunstatter1} and to remain
quasi-local we will be constrained in a box. The basic physical
requirement of quasi-locality that is reflected by the box
quantization can be also expressed as: any quantum spectra that would
heavily depend on the physics outside a constrained region are
necessarily unphysical, as the configurations of variables outside
that region are in some sense always arbitrary. Let us further mention
that the boundary conditions which are adapted at the walls of the box
already assume that one of the walls of the box is a spacetime
horizon. The constrain by the box means we have a limited domain in
the $x$ coordinate, with the walls of the box corresponding to
$x=x_{1}$ and $x=x_{2}$, where $x_{1}$ lies on some bifurcate horizon
$x_{1}=x_{H}$, and $x_{2}$ is just a fixed point in the static region
of spacetime.

The 2D metric in a suitable coordinates has only one degree of
freedom. The Birkhoff theorem holds for our case \cite{Kunstatter4}, which means
the solutions of our theory are \emph{static} and have a simple
structure. Let us therefore express the line element corresponding to
the solution in the static coordinates as:
\[ds^2_{_{2D}} = -F\cdot dT^{2}+F^{-1}d\phi^{2}.\]
$T$ is the Killing time (in the static region we are dealing with). The 
Birkhoff theorem for the solutions of the 2D dilaton-E-M theory further 
tells us that the function $F$ can be obtained \cite{Kunstatter4} as
\[F=I(\phi)-2M-Q^{2}K(\phi),\]
where 
\[I(\phi)=\int^{\phi}V(\phi')d\phi',\]
\[K(\phi)=\int^{\phi}\frac{d\phi'}{W(\phi')}.\]
 
$F$ can be expressed through the ADM data as:
\[F=\frac{\phi'^{2}}{B^{2}}-P_{B}^{2}.\]
Let us further define:
\[M\doteq\frac{1}{2}\left(-F+I(\phi)-P_{A_{x}}^{2}K(\phi)\right).\]
\[P_{A_{x}}\doteq Q.\] 

By connecting the physical parameters of the solution with the ADM data
one obtains the following relations \cite{Kuchar}
\begin{equation}\label{B}
B=\sqrt{-FT'^{2}+F^{-1}\phi'^{2}}
\end{equation}
\begin{equation}\label{v}
N^{x}=\frac{-F\dot T T'+F^{-1}\dot\phi \phi'}{-FT'^{2}+F^{-1}\phi'^{2}},
\end{equation}
\begin{equation}\label{u}
N=\frac{\phi'\dot T-T'\dot\phi}{\sqrt{-FT'^{2}+F^{-1}\phi'^{2}}}.
\end{equation}
Then plugging this in the expression for $P_{B}$ one can prove the following:
\[T'=-\frac{B P_{B}}{F}.\]
One can further show that by taking the Poisson brackets $T'$ commutes
with $Q$ and $\phi$ and behaves as a canonical conjugate to
$M$. Therefore let us further call it $P_{M}$ and suggest that after
some appropriate transformation of variables it will play role of a
canonical conjugate to $M$.  One can express the $P_{M}$ variable also
as:
\[P_{M}\doteq T'=-\frac{B^{3}P_{B}}{B^{2}P_{B}^{2}-\phi'^{2}}.\]

Now let us try to define a new canonical chart $\{M,P_{M},Q, P_{Q},
\phi, \tilde P_{\phi}\}$. This chart is much more relevant to capture
the physics of the problem we are interested in. In the new chart we
already know the variables $M,P_{M},Q,\phi$ and want to define the
remaining $P_{Q}$ and $\tilde P_{\phi}$. We proceed as in
\cite{Louko1, Louko2}, where the variables were derived particularly
for the Reissner-Nordstr\"om BH. Since $M, Q, \phi$ are all scalars,
the constraint variable $H_{x}$ has to be given as:
\begin{equation}\label{consvar}
H_{x}=P_{\phi}\phi'-\Lambda P_{\Lambda}'-A_{x}P_{A_{x}}'=P_{M}M'+P_{Q}Q'+\tilde P_{\phi}\phi',
\end{equation}
and finding the most natural solution of \eqref{consvar} one obtains:
\[P_{Q}=-\left(A_{x}+\frac{P_{A_{x}}K(\phi)B^{3}P_{B}}{B^{2}P_{B}^{2}-\phi'^{2}}\right)=-
A_{x}+P_{A_{x}}K(\phi)P_{M},\]
and
\begin{widetext}
\begin{eqnarray}
\tilde P_{\phi}=
P_{\phi}+\frac{\phi'(BP_{B})'-\phi''BP_{B}+\frac{1}{2}B^{3}P_{B}\{V(\phi)-P_{A_{x}}^{2}W^{-1}\}}{B^{2}P_{B}^{2}-\phi'^{2}}=~~~~~~~~~~~~~~~~~~~~~~~~~~~~~~~~~\\
P_{\phi}+\frac{\phi'(BP_{B})'-\phi''BP_{B}}{B^{2}P_{B}^{2}-\phi'^{2}}+\frac{1}{2}P_{M}\left(V(\phi)-\frac{P_{A_{x}}^{2}}{W(\phi)}\right).~\nonumber
\end{eqnarray}
\end{widetext}
It is important to note that these variables are generalizations of the variables obtained in 
Ref. \cite{Louko1}.

It is well known that the Hamiltonian constrains $H_{t}=H_{x}=H_{A}=0$
can be solved \cite{Kunstatter4} by:
\[M'=Q'=0,\]
together with the last condition that we get from the superhamiltonian
constraint
\[P_{M}M'+P_{Q}Q'+\tilde P_{\phi}\phi'=0,\]
and this is:
\[\tilde P_{\phi}=0.\]

Further, one needs to show that the transformation to the new variables $(M, Q, \phi)$ 
is canonical. One can generalize the result of Ref. \cite{Cavaglia1} as follows:
\begin{widetext}
\begin{eqnarray}
\int_{x_{1}}^{x_{2}}dx\left[\dot{M} P_{M}+\dot{\phi} \tilde P_{\phi}+\dot{Q} P_{Q}-\dot{B} P_{B}-\dot{\phi} P_{\phi}-\dot{A_{x}} P_{A_{x}}\right]
=\left(\int dx \cdot \dot{G}\right) + S(x_{2})-S(x_{1}),
\end{eqnarray}
\end{widetext}
where the function $G$ is:
\begin{widetext}
\[G=\left[\phi'\cdot\hbox{Arctanh}\left(\frac{\phi'}{FP_{M}}\right)-FP_{M}+Q\{P_{Q}-QK(\phi)P_{M}\}\right],\]
\end{widetext}
and the surface term function is
\[S(x)=\dot\phi\cdot\hbox{Arctanh}\left(\frac{\phi'}{P_{M}F}\right).\]
This surface term can be shown to vanish by requiring that $\dot\phi$
vanishes sufficiently quickly at the boundaries. This shows that the
transformation is canonical.

Our constraints show that the action \eqref{Gaction2} rewritten in the
new variables can be reduced by extremalizing the variables at each of
the hypersurfaces (or in other words by solving the constrains) to a
1-dimensional reduced action as:
\begin{equation}\label{reduced}
S_{red}=\int dt\cdot (\mathcal{P_{M}}\dot M+\mathcal{P_{Q}}\dot Q-H),
\end{equation}
where $H$ is some Hamiltonian to which also the surface terms of the
action contribute and
\[\mathcal{P_{M}}=\int_{x_{1}}^{x_{2}}P_{M}dx=T(x_{1})-T(x_{2}),\]
\[\mathcal{P_{Q}}=\int_{x_{1}}^{x_{2}}P_{Q}dx.\]

\subsection{Quantization in the Euclidean sector}


For a general horizon of the spherically symmetric static line element \eqref{static} 
there is a well defined concept of horizon temperature linked to the time 
periodicity of the regular solution in the Euclidean sector of the theory. Let us further argue as 
in Refs. \cite{Kunstatter1, Kunstatter2, Kunstatter3}:
Periodicity in $T$ in the Euclidean sector implies that $\mathcal{P_{M}}$ 
is also periodic with the same period $T_{H}^{-1}$. Imposing the periodicity condition on the action
\eqref{reduced} one can provide a transformation of variables:
\[ X=\sqrt{\frac{E(M)}{\pi}}\cos\left(2\pi \mathcal{P_{M}}T_{H}\right),\]
and
\[P_{X}=\sqrt{\frac{E(M)}{\pi}}\sin\left(2\pi \mathcal{P_{M}}T_{H}\right).\]
For this transformation to be canonical, as shown in \cite{Kunstatter1, Kunstatter2, Kunstatter3}, direct
calculation implies that the following must hold:
\begin{equation}\label{can}
\frac{\partial E(M)}{\partial M}=T_{H}^{-1}.
\end{equation} 
Now consider theory in which $M$ corresponds to some reasonably
defined mass, then \eqref{can} means $E$ is from the definition
related to the entropy of a horizon and some theory of gravity, in
particular
\[E(M)+S_{ext}(Q)=S_{H}.\]
Then one can easily observe:
\[S_{H}-S_{ext}(Q)=2\pi\left(\frac{1}{2}P_{X}^{2}+\frac{1}{2}X^{2}\right),\]
with the spectrum of the harmonic oscillator:
\begin{equation}\label{oscillator}
S_{H}=2\pi (n+1/2)+S_{ext}(Q).
\end{equation}   
This spectrum describes deviation of entropy from entropy of an
extremal state. 

\subsection{The theories reduced to the dilaton action}

Let us consider now a reduced theory to a 2D dilaton gravity. The 2D
reduced metric $g_{ab}^{red}$ of the theory generally relates to the
dilaton metric $g_{ab}$ via some conformal transformation:
\[g_{ab}^{red}=\Omega^{-2}g_{ab}.\]
The metric of the higher dimensional solution is therefore ($a,b=0,1$)
\begin{equation}\label{originalmetric}
ds^2 = \Omega^{-2}g_{ab}dx^{a}dx^{b}+r^{2}d\Omega_{D-2}.
\end{equation}
(The dilaton field $\phi$ from the action \eqref{Gaction} is a
particular function of $r$ that depends on the reduced theory.) The
horizon of the solution is defined as:
\[F(\phi_{H})=0 ~~\to~~M=\frac{1}{2}\left\{I(\phi_{H})-Q^{2}K(\phi_{H})\right\}.\] 

Now let us recover the thermodynamical concepts of the original metric
\eqref{originalmetric} that was reduced. The concepts of entropy and
temperature must depend beyond the dilaton theory also on the
reduction, which includes conformal transformation and redefinition of
variables. Taking care of this, one can easily observe that the
temperature of the horizon becomes:
\[T_{H}=\frac{\Omega^{-2}(\phi_{H})F_{,\phi|\phi_{H}}\phi_{,r~|\phi_{H}}}{4\pi}\]
and if we suppose that $M$ is the mass of the theory, the entropy of the horizon turns out to be ($dM=\frac{1}{2}F_{,\phi|\phi_{H}}d\phi_{H}$):
\begin{equation}\label{entropy2}
S_{H}=\int T_{H}^{-1}dM=\int \frac{2\pi\Omega^{2}}{\phi_{,r|\phi_{H}}}d\phi_{H}=\int 2\pi\Omega^{2}(r_{H})dr_{H}.
\end{equation}

The above expression raises the following two important questions: In the spherically symmetric sector, which gravity theories  
reduce to the 2D dilaton theory? Also, what kind of quantization spectra will this dilaton gravity give for Wald entropy? 
As expected $D>2$ dimensional GR with cosmological constant is included in our picture,
and $M$ represents the ``correct'' mass. In particular for GR in $D$
dimensions with cosmological constant we can write the functions
$\Omega^{2}$, $V(\phi)$, $W(\phi)$ and $\phi(r)$ from the action
\eqref{Gaction} as follows:
\[V(\phi)=2(D-3)\left(\frac{\alpha\phi}{2}\right)^{-\frac{1}{D-2}}-4\Lambda\left(\frac{\phi}{2}\right)^{\frac{1}{D-2}},\]
\[W(\phi)=(D-2)\alpha^{\frac{D-3}{D-2}}\left(\frac{\phi}{2}\right)^{\frac{2D-5}{D-2}},\]
\[\phi(r)=\alpha^{-1}r^{D-2},\] 
\[\Omega^{2}(r_{H})=2(D-2)\alpha^{-1}r^{D-3}.\]
Here we define the symbol $\alpha$ to be $\alpha=16\pi/V_{D-2}$ with
$V_{D-2}$ being the volume of a unit $D-2$ sphere.  Then the upper
expression for entropy \eqref{entropy2} gives in the case of GR in $D$
dimensions:
\[S_{H}=\frac{V_{D-2}r_{H}^{D-2}}{4}=\frac{A_{H}}{4}.\]

The Lovelock action can be reduced in the spherically symmetric sector to
\cite{Cvitan}:
\begin{widetext}
\[S =V_{D-2}\int dtdx ~\sum_{m=0}^{[D/2]}\left\{\lambda_{m}\frac{(D-2)!}{(D-2m)!}mR^{(2)}r^{2}[1-(\triangledown r)^{2}]+\tilde L_{m}(r,\triangledown r,\triangledown^{2}r)\right\},\]
\end{widetext}
where $\tilde L_{m}$ is some complicated highly non-linear function of the
covariant derivatives of $r$. However, near the horizon one can apply
the condition of $(\triangledown r)^{2}$ being small, and neglect any higher powers
of $(\triangledown r)^{2}$, simplifying $\tilde L$ in a way that one obtains
\cite{Cvitan}, (after suitable redefinitions of variables and a
conformal transformation), the action \eqref{Gaction}. The near
horizon approximation means that we take the approximation of the box
in which we quantize not extending ``too far'' from the horizon\footnote{However, it is nice to see that the entropy spectra \eqref{oscillator} can be obtained for the 5D Gauss-Bonnet BH without this approximation by using the geometrodynamical approach from \cite{Simon}.}.

The conformal transformation is such that the function $\Omega^{2}$ is
given as $\Omega^{2}=d(\Phi^{2})/dr$ with $\Phi$ being a function of
$r$ as (but the exact form of the function actually turns not to be
important for the consequence we are trying to make):
\[\Phi^{2}=2V_{D-2}\sum_{m}^{[\frac{D}{2}]}\left(m\lambda_{m}\frac{(D-2)!}{(D-2m)!}r^{D-2m}\right).\]
The one can easily observe from \eqref{entropy2}:
\[S_{H}=2\pi\Phi_{H}^{2}.\]
This is precisely the result for the entropy in the Lovelock gravity
\cite{Cvitan}. This means that in the near horizon approximation the
previous result contains also arbitrary Lovelock gravity in $D$
dimensions. 


\subsection{Discussion of the results}

In the previous section we obtained the results for 2D dilaton gravity with aribtrary 
potential and presence of E-M ield. First of all in papers
\cite{Kunstatter1, Kunstatter2, Kunstatter3} the ideas were applied to the BH horizons. However since there
is nothing that in the previous analysis distinguishes the BH horizon
from other horizons, we suggest here that if a static spherically
symmetric spacetime is a solution of GR, the quantization as described
above can be applied in the same way to the entropy (or more specifically to the
deviation of entropy from the extremality) of \emph{any arbitrary}
horizon of the line element \eqref{static}. Therefore, as mentioned before, (since
cosmological constant is included) we can apply the quantization from \cite{Kunstatter1, Kunstatter2, Kunstatter3} 
to quantize entropy of the deSitter cosmological horizon, or even (considering
the interior region inside the Cauchy horizon), to the inner Cauchy
horizon of the R-N black hole. Furthermore it was shown how the result concerning
quantization of entropy can be (as some kind of asymptotic result)
generalized to the entropy of the horizons within Lovelock
gravity. Therefore we claim that the results of \cite{Kunstatter1,
  Kunstatter2, Kunstatter3}, when pushed to their consequences,
support the basic conclusion of this paper.

However, let us point the fact that we trust the spectra \eqref{oscillator}
only in the semi-classical limit. This is due to the fact that the
calculation from \cite{Kunstatter1, Kunstatter2, Kunstatter3} is
essentially a semi-classical calculation, combining classical and
quantum ideas. (In order to be able to exactly quantize the theory one
has to freeze most of the degrees of freedom by requiring spherical
symmetry and more importantly, by reducing the action to one dimension
by plugging into the action the \emph{classical} solutions of the
theory.)  The fact that the above spectra are valid \emph{only} in the 
semi-classical limit implies that the spectra may not be of the form 
of Bekenstein and Mukhanov \cite{Mukhanov}:
\[\delta S_{H}=\ln(k), ~~k\in\mathbb{N}_{+}.\]  

At the end of this section let us discuss one more problem: Like 
in Refs. \cite{Kunstatter1, Kunstatter2, Kunstatter3}, let us identify the
$S_{ext}(Q)$ function in the spectrum \eqref{oscillator} with the
entropy of the extremal case.  The $S_{ext}(Q)$ function can be then
identified with some classical expression for the entropy of the
extremal spacetime as a function of charge. (For the
Reissner-Nordstr\"om spacetime it is for example $S_{ext}=\pi Q^{2}$.)
If the charge has to be quantized in integer multiples of an
elementary charge, one might wonder if this condition does not
contradict the general entropy quantization rule for each of the
spacetime horizons. The somewhat naive calculation shows, that even in
the Reissner-Nordstr\"om spacetime one can derive from the spectrum
\eqref{oscillator} (plugging in the correct quantization of charge) an
area/entropy spectrum of the inner Cauchy horizon and it will
\emph{not} be equispaced.

However we want to point one significant insufficiency of the reasoning
that leads to this apparent contradiction. Certainly every classical
spacetime has to be \emph{quasi-locally} approximated by a quantum
spacetime in the semi-classical limit (Bohr's principle of
correspondence). However one cannot apply this reasoning to the spacetime
as a whole. Take a simple example of the Schwarzschild spacetime:
despite of the fact that quasi-locally (in limited regions) it
certainly has to be approached by some quantum configuration in the
semi-classical limit, \emph{globally} this does not hold. Unlike the
classical Schwarzschild spacetime which is static, any quantum
configuration would Hawking radiate and on sufficiently large time
scale it would rapidly deviate from the classical static solution. In
Reissner-Nordstr\"om spacetime the deviation on the semi-classical
level would be even faster: The inner Cauchy horizon would be
relatively rapidly destroyed by a mass-inflation instability triggered
by a backscattered Hawking radiation coming from the horizon itself.

Therefore one cannot claim that a classical $n$-parameter family of
solutions has to be \emph{globally} approached by a corresponding
quantum family with semi-classically quantized parameters. Any of the
results we presented is for fundamental reasons \emph{quasi-local} and
so is the classical limit. Accepting the quasi-locality of the result
removes the above apparent contradiction and all the horizon spectra
could easily co-exist.

\medskip

\section{Conclusions}

In this work we have generalized different results typically derived for the 
black hole horizons in GR, to general horizons in generalized gravity theories. 
We showed that all three standard arguments of the entropy spectra --- Bekenstein's
universal lower bound on the entropy transition, the highly damped quasi-normal modes 
and the reduced phase-space quantization --- can be generalized, and imply the following: The entropy 
quantization of the Bekenstein form \eqref{entropy} is a \emph{robust} result and holds 
for a generic horizon (at least in the static spherically symmetric sector) of a wide class 
of gravity theories. Furthermore the results lead to the choice of $\gamma=1$ in the Bekenstein 
type spectra \eqref{entropy}.

Similarly to Ref. \cite{Paddy1} we have shown that in the theories where proportionality between 
horizon area and horizon entropy does not hold, entropy is the quantity that remains equispaced. 
The entropy spectra therefore provide another example of a result, such that was originally derived within
the black hole thermodynamics, but generalizes to the horizon thermodynamics in reasonably general 
gravity theories.

\medskip

\section*{Acknowledgments}

The work is supported by Max Planck-India Partner Group on Gravity and
Cosmology. SS is partially supported by Ramanujan Fellowship of DST,
India.

\end{document}